\documentclass[journal]{IEEEtran}
\usepackage[pdftex]{graphicx}
\usepackage{subfigure}
\usepackage{balance}
\usepackage{array}
\usepackage{times}
\usepackage{rotating}
\usepackage{multirow}
\usepackage[cmex10]{amsmath}
\usepackage{subfig}
\usepackage{tabularx}
\usepackage{float}
\usepackage{breqn}
\usepackage{siunitx}
\usepackage{amsfonts}
\usepackage[mathscr]{euscript}
\usepackage{epstopdf}
\usepackage{amsmath}
\usepackage{amssymb}
\usepackage{fancyhdr}
\newenvironment{definition}[1][Definition]{\begin{trivlist}
\item[\hskip \labelsep {\bfseries #1}]}{\end{trivlist}}

  \newcommand{\bs}{\mathbf{s}}

  \newcommand{\bw}{\mathbf{w}}

  \newcommand{\bx}{\mathbf{x}}

  \newcommand{\bF}{\mathbf{F}}

  \newcommand{\bH}{\mathbf{H}}

  \newcommand{\bU}{\mathbf{U}}

\newtheorem{lemma}{Lemma}

\newtheorem{theorem}{Theorem}

\newcommand{\beq}{\begin{equation}}

\newcommand{\enq}{\end{equation}}

\newcommand{\beqa}{\begin{eqnarray}}

\newcommand{\enqa}{\end{eqnarray}}

\newcommand{\bea}{\begin{array}}

\newcommand{\ena}{\end{array}}

\newcommand{\bef}{\begin{figure}}

\newcommand{\enf}{\end{figure}}

\newcommand{\bds}{\begin {itemize}}

\newcommand{\eds}{\end {itemize}}

\newcommand{\bdf}{\begin{definition}}

\newcommand{\blm}{\begin{lemma}}

\newcommand{\edf}{\end{definition}}

\newcommand{\elm}{\end{lemma}}

\newcommand{\bthm}{\begin{theorem}}

\newcommand{\ethm}{\end{theorem}}

\newcommand{\cI}{{\ensuremath{\mathcal{I}}}}

\DeclareMathAlphabet{\mathcalligra}{T1}{calligra}{m}{n}
\begin{document}

\title{Pulse Shaping Diversity to Enhance Throughput in Ultra-Dense Small Cell Networks}

\author{Amir H.~Jafari{$^{12}$}, Vijay Venkateswaran{$^{3}$}, David~L\'opez-P\'erez{$^{2}$}, Jie~Zhang{$^{1}$}\\

\
{$~^{1}$}Dept. of Electronic \& Electrical Engineering, University of Sheffield, Sheffield S1 3JD, UK\\
{$~^{2}$}Bell Laboratories Nokia, Dublin, Ireland\\
{$~^{3}$}Huawei Technologies, Sweden\\}
\maketitle

\maketitle

\begin{abstract}
Spatial multiplexing (SM) gains  in multiple input multiple output (MIMO) cellular networks are limited when used in combination with ultra-dense small cell networks. 
This limitation is due to large spatial correlation among channel pairs. 
More specifically, it is due to i) line-of-sight (LOS) communication between user equipment (UE) and base station (BS) 
and ii) in-sufficient spacing between antenna elements. 
We propose to shape transmit signals at adjacent antennas with distinct interpolating filters which introduces pulse shaping diversity eventually leading to improved SINR and throughput at the UEs. In this technique, each antenna transmits its own data stream with a relative offset with respect to adjacent antenna. The delay which must be a fraction of symbol period is interpolated with the pulse shaped signal and generates a virtual MIMO channel that leads to improved diversity and SINR at the receiver. Note that non-integral sampling periods with inter-symbol interference (ISI) should be mitigated at the receiver. For this, we propose to use a fractionally spaced equalizer (FSE) designed based on the minimum mean squared error (MMSE) criterion.
Simulation results show that for a $2\times2$ MIMO and with inter-site-distance (ISD) of 50~m, the median received SINR and throughput at the UE improves by a factor of 11~dB and 2x, respectively, which verifies that pulse shaping can overcome poor SM gains in ultra-dense small cell networks. 

\end{abstract}

\section{Introduction}

One of the most promising approaches to meet the data deluge and to enhance network capacity is small cell densification, 
which benefits from extensive spatial reuse of the spectrum~\cite{2015Lopez}. 
Having multiple antennas and exploiting MIMO technology can further enhance the network capacity~\cite{1532224}~\cite{LTE}. Applying MIMO tools such as spatial multiplexing in 
ultra-dense small cell networks introduces new challenges due to different propagation conditions when compared to macro cell scenarios. For instance, spatial multiplexing (SM) gain typically improves when the MIMO spatial channels are uncorrelated. Ultra-dense small cell networks are limited by channels being correlated and SM gains are very limited. Two phenomena contribute to this channel correlation. Firstly, antennas at both user equipment (UE) and base station (BS) are placed very closely to each other ($\sim$ half wavelength). Secondly, due to proximity of both UE and BS, there is a high probability of LOS communication resulting in high spatial correlation. Thus the communication channel tends to be ill-conditioned, lowering the number of independent parallel data streams that can be simultaneously multiplexed and transmitted and, therefore, the throughout is considerably degraded. 

\subsubsection*{\textbf{Central Idea}}
In order to improve throughput in ultra-dense small cell networks, 
we propose a novel transmission technique using distinct pulse shapes to modulate adjacent antennas' data streams. We refer to this technique as diversity pulse shaped transmission (DPST). 

One way to view DPST is that adjacent antenna element signals are shaped with slightly different band limited pulse shaping filters. This change will introduce delay and ISI in time domain. This delay which must be a \emph{fraction of symbol period} allows the UE receiver to sense multiple delayed replicas of the transmitted data stream. Consider a $2\times2$ MIMO setup for simplicity, 
the receive antennas would observe antenna~1 transmitting its data stream with symbol period $T_s$, 
as well as antenna~2 transmitting its data stream; however, with delay $\tau$ with respect to antenna~1 ($0 < \tau < T_{s}$) while being sampled at $T_s$. This implies that in a LOS scenario, a receiver that is synchronized to symbol period $T_s$ (ignoring the bulk delays) would observe a direct path with data stream~1 as well as a delayed path with data stream~2, with the latter corrupted by inter-symbol interference (ISI).

The improved diversity is somewhat inspired from faster than Nyquist (FTN) signaling~\cite{6479673}~\cite{4777625}~\cite{1231648}. In FTN, data streams are sampled and transmitted at a fraction of symbol period, eventually leading to an improvement in communication rates at the cost of complicated receiver to combat ISI. In our case, to account for the increased ISI in the system, we propose to use a fractionally spaced equalizer (FSE) at the UE that operates on the precoded data streams and wireless channel output and eventually leads to improved diversity gain in ultra-dense small cell networks~\cite{489269}. In order to ensure a reasonable estimate of multipath signals, minimum mean squared error (MMSE) criterion is used to design the equalizer. Indeed, DPST can also be seen as an application of FTN and FSE to enhance data rates. 

\subsubsection*{\textbf{Contributions}}

In this paper, we present DPST whose arrangement based on pulse shaping diversity and oversampled receiver, improves the overall dimensionality of the transmitted multi-antenna data streams viewed at the receiver and consequently enhance the data rates. 
Moreover, we evaluate the performance of DPST in a single tier hexagonal multi small cell layout. In more detail, we quantify the degradation of SINR due to spatial correlation for a $2\times2$ MIMO system at different inter-site-distances (ISDs), i.e., 20~m, 50~m and 150~m, and show that the proposed DPST leads to a 50\%-tile SINR improvement of 11~dB at an ISD of 50~m. We also show that DPST can almost enhance the UE throughput by 2x at all ISDs.

It is important to highlight that DPST differs from cyclic delay diversity (CDD)~\cite{4656965}~\cite{989781}~\cite{4109390}, a diversity technique used in LTE.  
In CDD, the antennas transmit a cyclically shifted version of the same signal to achieve diversity gain, whereas in DPST each antenna transmits an individual signal with a fractional delay with respect to its former antenna which requires to redesign the precoder as well as the receiver in order to enhance spatial multiplexing gains. 
We also feel it is necessary to note that while exploiting antenna polarization at the transmitter is an effective technique in MIMO systems, it is usually limited to two transmit antennas~\cite{1192168}. 
In contrast, DPST can be applied to larger MIMO systems subject to derivation of the optimized fractional amount that is applied to transmission of each transmit antenna.

The rest of this paper is as follows. Section~\ref{sec:part_DPST} details the proposed DPST. Section~\ref{sec:DPST_rec} explains the DPST receiver design, precoder and equalizer to estimate the transmitted signals. 
Section~\ref{sec:part_Simulation} presents a performance evaluation/comparison of DPST with existing MIMO systems. Section~\ref{sec:part_Conclusion} draws the conclusions.

\begin{figure*}[t]
\centering
\includegraphics[scale=0.55]{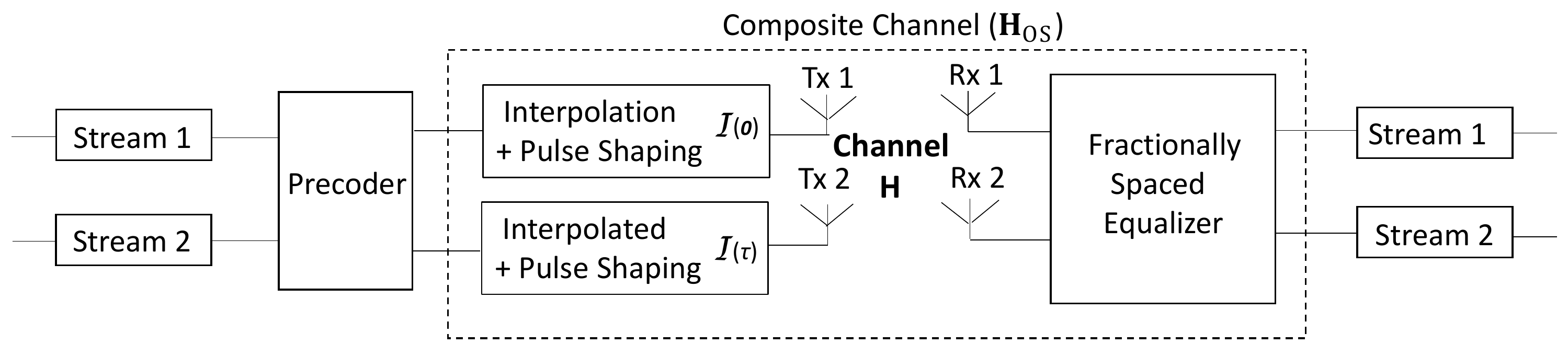}
\caption{DPST block diagram.}
\label{fig:DPST_Block}
\vspace{-0.3cm}
\end{figure*}

\section{Diversity Pulsed Shaped Transmission}
\label{sec:part_DPST}
\subsection{Channel Correlation Model}

The MIMO performance is highly dependent on the overall channel taps being spatially uncorrelated.
In correlated small cell channel scenarios, different transmit-receive antenna pairs will experience similar channel conditions, and as a result the multipath components corresponding to different pairs may not be resolvable by the UE~\cite{1203167}. Thus correlated channels have reduced degrees of freedom, leading to reduced throughputs as detailed in~\cite{1459054}~\cite{892194}.

The Rician multipath fading model used to capture the LOS communication can be used to represent the MIMO channel as
\begin{equation}
  \textbf{H} = \sqrt{\frac{K}{K+1}} {\textbf{H}_{i}}^{\rm LOS} + \sqrt{\frac{1}{K+1}}{\textbf{H}_{s}}^{\rm NLOS},
  \label{eq:LOS_NLOS}
\end{equation}
where $\rm K$ is the Rician K factor~\cite{2015Jafari} modeled as a function of the UE-BS distance \textit{d},
i.e.,
\begin{equation}
  \rm K=\begin{cases}
      32 & \text{if $d<18 m$} \\
     140.10 \times \exp(-0.107 \times d) & \text{otherwise},
  \end{cases}
  \label{eq:Kfactor}
\end{equation}
and ${\textbf{H}_{i}}^{\rm LOS}$ and ${\textbf{H}_{s}}^{\rm NLOS}$ are the identity LOS MIMO channel matrix and the correlated non-LOS (NLOS) MIMO channel matrix, respectively. 

It is important to note that condition number ($\mathcal{K}$) is a metric to denote the number of independent streams of the wireless channel, and corresponds to the ratio of the maximum to minimum singular values of the wireless channel $\bH$~\cite{Matrix}. 
$\mathcal{K} \approx 1$ implies no correlation between channel pairs, 
and as long as $\mathcal{K}$ is less than $10$, the channel is regarded as well-conditioned, 
and can be leveraged to extract the unitary vectors of the channel for precoding and spatial multiplexing purposes. 
In ultra-dense small cell networks,
$\mathcal{K}$ is considerably larger than 10~\cite{2015Jafari}, 
and thus the channel becomes ill-conditioned and spatial multiplexing gain suffers.

\subsection{MIMO Link Model with Pulse Shaping}
\label{ssec:Transmitter}

For simplicity, we will model $2\times2$ MIMO channel used with DPST. Note that, DPST can be applied to arbitrary sized arrays. In the $2\times2$ MIMO system, DPST works by shaping the data streams of the second antenna with respect to the first antenna such that in downlink communications, the multipath components from second antenna arrive at the UE with a time offset when compared to that of the first antenna.
This enhances diversity in \textit{fractionally delayed} multipath components of the closely placed transmit antennas in a MIMO system, and allows to improve the UE throughput. Fig.~\ref{fig:DPST_Block} provides a block diagram of the proposed setup showing pulse shaping at the transmit side. 

\begin{figure}[b]
\centering
\includegraphics[scale=0.58]{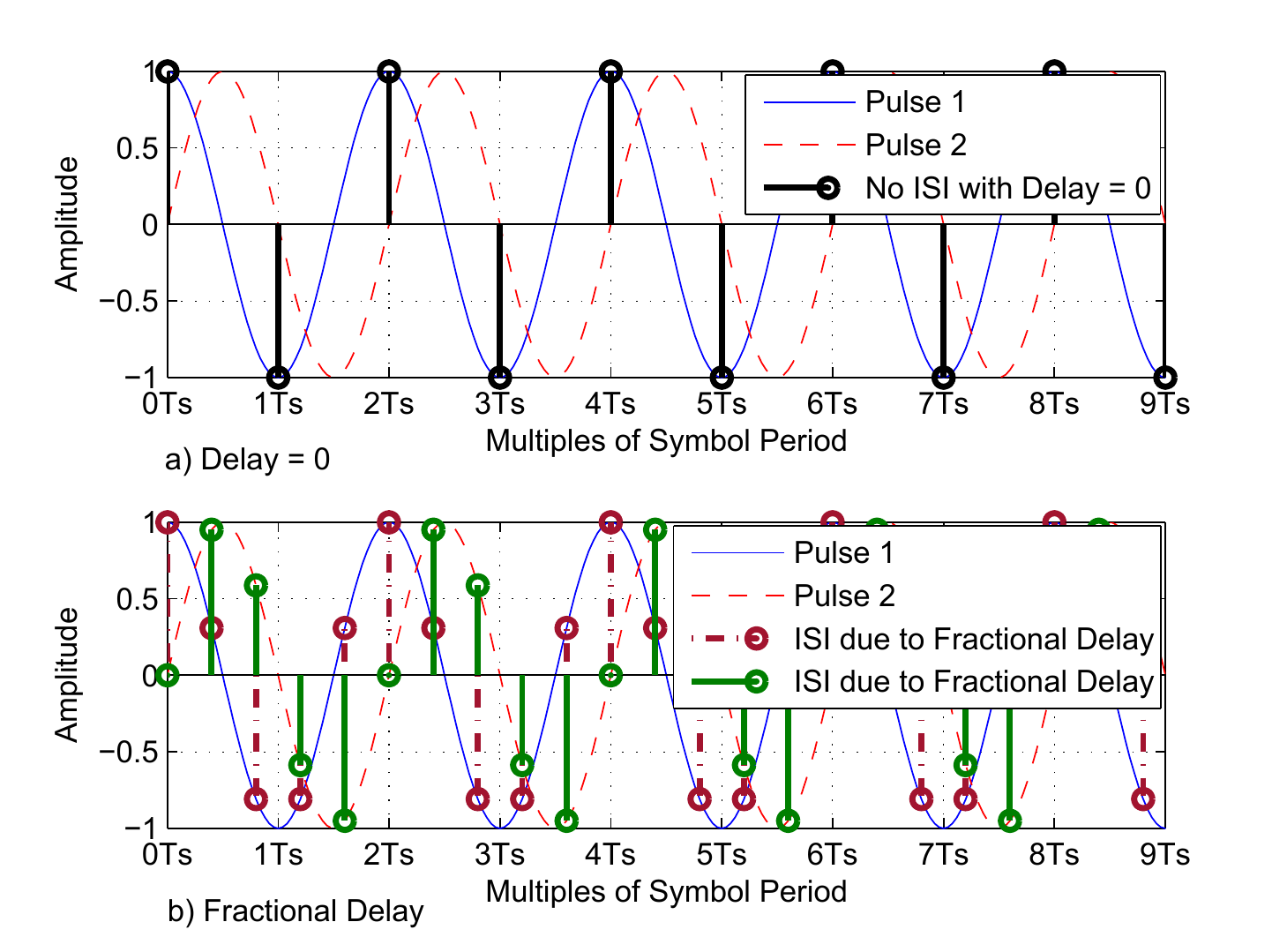}
\caption{Analogous performance of DPST and FTN.}
\label{fig:FTN_DPST}
\vspace{-0.3cm}
\end{figure}

The fractionally delayed signaling can be related to FTN communication, where a non-orthogonal sampling kernel is used to allow for signaling above the Nyquist limit while introducing ISI. In traditional cases, we typically use orthogonal sampling kernel such as a sinc pulse where an integral multiple of the symbol period contains only a single non-zero component of the transmitted signal. 
In FTN shaping, the signals operate above the Nyquist rate and the ISI that is intentionally introduced through oversampling results in enhanced system capacity~\cite{6479673}~\cite{4777625}~\cite{1231648}. Along similar lines, the \textit{fractional} delay imposed on the transmission of the second antenna in DPST injects additional \textit{deterministic} ISI to the system, which suggests the analogy between the performances of DPST and FTN. Fig.~\ref{fig:FTN_DPST} intuitively shows the implication of DPST. Fig.~\ref{fig:FTN_DPST}a shows signals sampled at \textit{integral} multiples of symbol period with an orthogonal sampling kernel, while Fig.~\ref{fig:FTN_DPST}b shows sampling at \textit{non-integral} multiples of symbol period. Note that in the latter case and in contrast with the former, at each sampling instant, there are multiple non-zero samples viewed by the sampling kernel. As a result, this can be exploited to increase the diversity of the wireless channel seen between the closely placed transmit and receive antennas. 

When viewed from the receiver perspective, the pulse shaping operation at the transmitter can be perceived as a fractionally delayed transmission with respect to the first transmit antenna where the delay $\tau$ is a fraction of the transmitted symbol period $T_s$, 
i.e., $0 < \tau < T_s$, and $T_s$ is assumed to be~$1$.
The received signal can be expressed as
\[
\bea{ccc}
\left[\bea{c}
x_{1}(t) \\ x_{2}(t)
\ena\right]
& = &
\left[\bea{cc}
h_{1,1}(t) & h_{1,2}(t) 
\\ h_{2,1}(t) & h_{2,2}(t)
\ena\right]
\ast
\left[\bea{c}
s_{1}(t) \\ s_{2}(t+\tau)
\ena\right]
\vspace{0.2cm}
\\
\bx(t) & = & \bH(t) \ast \bs(t)
\ena
\]
where $\textbf{H}(t)$ represents a continuous time version of the $2\times2$ correlated MIMO channel $\bH$, 
$s_{1}(t)$ and $s_{2}(t)$ are the signals transmitted by first and second transmit antennas, respectively, 
and $\ast$ is the convolution operation.
In discrete time domain, the delay $\tau$ imposed to the transmission of the second antenna is alternatively modeled by oversampling/interpolating the transmit data streams as shown in the following

\beqa
\left[\bea{@{}c@{}}
x_{1}(t) 
\\ x_{2}(t) 
\ena\right]
=
\left[ \bea{@{}cc@{}}
h_{1,1}(t) & h_{1,2}(t) \\ 
h_{2,1}(t) & h_{2,2}(t) 
\ena\right] 
\ast
\left(
\left[\bea{@{}cc@{}}
\cI(0) & 0 
\\ 0 & \cI(\tau)
\ena\right]
\left[\bea{@{}c@{}}
s_{1}(t) \\ 
s_{2}(t) 
\ena \right]
\right)
\label{eq:21}
\enqa
where $\textbf{\ensuremath{\mathcal{I}}}(0)$ and $\textbf{\ensuremath{\mathcal{I}}}(\tau)$ are interpolation kernels that are applied to the first and second transmit antennas, respectively. The elements of $N\times M$ interpolation matrix are obtained as
\begin{equation}
\ensuremath{\mathcal{I}}_{nm} = sinc (\frac{n(\frac{T_{s}}{N})+ \tau -m(\frac{T_{s}}{M})}{\frac{T_{s}}{M}}) \quad
\begin{cases}
\textit{m} = 1,2,...,M \\ \textit{n} = 1,2,...,N 
\end{cases}
\label{eq:22}
\end{equation}
where $M$ is the number of input signal samples and $N = M R$ with $R$ being the oversampling ratio~\cite{5706377}. It is evident that the interpolation matrix corresponding to antenna~1 has no time offset, while the one corresponding to antenna~2 is offset by time $\tau$ to account for the delayed pulse shaping. For more discussion on the modeling of oversampled analog signals refer to~\cite{5706377}. Also, note that the transmitted signals in antenna~1 and antenna~2 are not matched anymore, and thus the corresponding signal degradation is taken into account.

The delay $\tau$ plays a key role in performance of DPST. A delay of $\tau=0$, results in expression (\ref{eq:21}) to collapse into the ill-conditioned wireless channel $\bH$, and does not improve diversity. For a delay that is an integral multiple of symbol period $\tau=k  T_s$ ($k = 1,2,...$), the integral DPST provides a cyclic shift of transmitted data streams at the receiver. We can interpret this integral delay as a variation of CDD~\cite{4656965}~\cite{989781}~\cite{4109390} which does not lead to significant improvements in throughput. However, when the delay is a fraction of symbol period $\tau \neq k  T_s$ ($k = 1,2,...$), the matrix $\ensuremath{\mathcal{I}}(\tau)$ becomes a block diagonal matrix operating on streams of the input sequence $\{s_2[1], \, s_2[2] \, \cdots s_2[k]\}$. 
The latter scenario is a consequence of using distinct pulses which generates deterministic ISI, thereby increasing the order of the composite channel between the transmit and receive streams as shown in Fig.~\ref{fig:DPST_Block}. Indeed, the signals corresponding to adjacent antenna elements are shaped with pulse shaping filters having different bandwidths. The change in bandwidth will introduce a delay and hence ISI in time domain.

From the receiver perspective, the signals from the two transmit antennas are observed at $kT_{s}$ and $kT_{s}+\tau$ ($k = 1,2,...$). However, this delay can only be observed and the induced diversity can only be extracted when the receiver is operating at a rate much greater than the symbol rate. The next section discusses the DPST receiver design.

\section{Receiver and Precoder design}
\label{sec:DPST_rec}

\subsection{Receiver Design Considerations}
\label{ssec:Receiver}

The sinc interpolation shown in ~(\ref{eq:21}) can be interchanged and be alternatively represented as

\beqa
\left[\bea{@{}c@{}}
x_{1}(t) 
\\ x_{2}(t) 
\ena\right]
=
\left( 
\left[ \bea{@{}cc@{}}
h_{1,1}(t) & h_{1,2}(t) \\ 
h_{2,1}(t) & h_{2,2}(t) 
\ena\right] 
\ast
\left[\bea{@{}c@{}c@{}}
\cI(0) & 0 
\\ 0 & \cI(\tau)
\ena\right]
\right)
\left[\bea{@{}c@{}}
s_{1}(t) \\ 
s_{2}(t) 
\ena \right]
\nonumber
\enqa

Interference is the main challenge in ultra-dense small cell networks (interference-limited) and, therefore, ignoring the noise term, the received signal at antenna 1 can be rewritten  as
\[
x_{1}(t) = 
\left[ \bea{cc}
\cI(0) \ast h_{1,1}(t) & \cI(\tau) \ast h_{1,2}(t)
\ena\right] 
\left[\bea{c}
s_{1}(t) \\ 
s_{2}(t) 
\ena \right]
\]

At the receiver side, DPST requires the receiver to operate at a rate significantly greater than the symbol rate. This is done by sampling the received signals $P$ times (typically $P\geq 2$) within the time interval $t \in \left(kT_s, \, [k+1]T_s\right]$ where $t = (k+\frac{1}{P})T_s$ and is stacked as a $P\times1$ vector, i.e.,
\[
\bea{ccc}
\left[ \bea{@{}c@{}}
x_{1}(t=kT_s) \\ \vdots \\ x_{1}(t=(k+\frac{P-1}{P})T_s)
\ena\right]
= \
\cI_{\rm R}
\left[\bea{@{}cc@{}}
\cI(0) \ast h_{1,1}(t) \\ \cI(\tau) \ast h_{1,2}(t)
\ena\right]^{T}
\left[\bea{@{}cc@{}}
s_{1}(t) \\ 
s_{2}(t) 
\ena \right]
\vspace{0.3cm}
\ena
\]

\beqa
\bx_{1,os}[k] & = & \bH_{1,os}[k] \ast \bs[k]
\nonumber
\enqa
where $\textbf{s}[k] = \textbf{s}(t = k T_{s})$ and $\textbf{x}_{1,os}$ is the interpolated received signal at the first antenna and is represented as a $P \times 1$ vector. $\cI_{R}$ is the $(N \times P) \times N$ sinc interpolation matrix similar to (\ref{eq:22}) exploited at each receive antenna and $()^{T}$ denotes the transpose. Note that operating at $P$ times the symbol rate at the receiver allows the UE to suppress the ISI using an equalizer. 

The received signal at antenna 2 is also sampled $P$ times and, therefore, stacking the $P \times 1$ vectors $\textbf{x}_{1,os}[k]$ and $\textbf{x}_{2,os}[k]$ from both receive antennas, we get

\beq
\bx_{os}[k] =
\left[ \bea{c}
\bx_{1,os}[k]\ \\ \bx_{2,os}[k]
\ena\right]
=
\left[ \bea{c}
\bH_{1,os} \\ \bH_{2,os}
\ena\right]
\ast
\bs[k]
= 
\bH_{os} \ast \bs[k]
\label{eq:27}
\enq

It is realized from~(\ref{eq:27}) that the resultant channel matrix $\textbf{H}_{\rm os}$ is a tall matrix, which can be a full rank column matrix in case of sufficient pulse diversity. Indeed, due to induced pulse diversity and antenna array phase shifts, the columns of $\textbf{H}_{1,\rm {os}}$ and $\textbf{H}_{2,\rm {os}}$ are independent of each other. This implies that the rank of the composite channel $\textbf{H}_{\rm os}$ is greater than 1, in contrary to the initial correlated channel $\textbf{H}$ with rank 1. This rank improvement is the outcome of
\begin{itemize}
\item the introduction of distinct pulse shapes and deterministic fractional delay, and 
\item higher order channel observed at the receiver due to oversampling. 
\end{itemize}
Moreover, it has to be noted that while SINR at a given sampling instant can be degraded due to pulses interfering with each other, 
the overall SINR after receiver equalization is still enhanced by exploiting the pulse diversity. 
 
From the receiver point of view, DPST enhances the channel degrees of freedom by virtually reducing the correlation between channel pairs. The overall channel $\textbf{H}_{\rm os}$ can be compactly represented as
\[
\textbf{H}_{\rm os} = \textbf{\ensuremath{\mathcal{I}}}_{\rm Rx} (\textbf{H} \ast \textbf{\ensuremath{\mathcal{I}}}_{\rm Tx}),
\label{eq:28}
\]
where $\textbf{\ensuremath{\mathcal{I}}}_{\rm Tx}$ and $\textbf{\ensuremath{\mathcal{I}}}_{\rm Rx}$ consist of the interpolation matrices for both antennas at the transmitter and receiver, respectively. To downsize $\textbf{H}_{\rm os}$ with respect to $\textbf{H}$, 
the channel is decomposed using singular value decomposition (SVD) as $\textbf{H}_{\rm os} = \textbf{U}_{\rm os} $\mbox{\boldmath{$\Sigma$}}$_{\rm os} \textbf{V}_{\rm os}^{\rm {H}}$ 
and since $\textbf{H}$ is a $2\times2$ matrix, 
$\textbf{U}_{\rm os}$ and $\textbf{V}_{\rm os}$ only take the first two columns and rows. Therefore, the downsized channel is 
\[
\textbf{H}_{\rm R} = \textbf{U}_{\rm os}(:,1:2)^{\rm {H}} \ \textbf{H}_{\rm os} \ \textbf{V}_{\rm os}(1:2,:)^{\rm {H}}
\]
The downsized channel $\textbf{H}_{\rm R}$ is also normalized with respect to $\textbf{H}$ as
\[
\textbf{H}_{\rm N} = \textbf{H}_{\rm R} \times \frac{\parallel \textbf{H} \parallel_{\rm F}}{\parallel \textbf{H}_{\rm R} \parallel_{\rm F}}.
\]
where $\parallel$ $\parallel_{\rm F}$ denotes the Frobenius norm.

\subsection{Precoding and Detection}
\label{ssec:prec_detec}

As previously discussed, the virtual channel $\textbf{H}_{\rm N}$ due to introduction of DPST is a channel with enhanced diversity. 
To perform closed loop precoding at the transmitter, 
we do SVD and extract the first two columns of $\textbf{V}_{\rm N}$ from $\textbf{H}_{\rm N} = \textbf{U}_{\rm N} $\mbox{\boldmath{$\Sigma$}}$_{\rm N} \textbf{V}_{\rm N}^{\rm {H}}$ to generate the precoding matrix denoted by $\textbf{W}$ where $\textbf{W} = \textbf{V}_{\rm N}(:,1:2)$. The precoding matrix is then subject to power scaling as the precoding must not violate the BS transmission power constraint, i.e.,
\[
\textbf{W} = \sqrt{P_{ \rm BS}} \times \rho \times \textbf{W},
\]
where $P_{\rm BS}$ is the BS transmit power and $\rho$ is the power scaling ratio. 
The precoded channel $\textbf{H}_{\rm eq}$ is then defined as $\textbf{H}_{\rm eq} = \textbf{H}_{\rm N} \textbf{W}$.

At the receiver side, minimum mean squared error (MMSE) equalizer is exploited and hence the receive filter is formulated from $\textbf{H}_{\rm eq}$ as
\[
\bea{ccc}
\textbf{F}^{\rm {MMSE}} & =& \textbf{H}_{\rm eq}^{\rm {H}}(\textbf{H}_{\rm eq}\textbf{H}_{\rm eq}^{\rm {H}} + \textbf{$\Phi$} + \textit{N}_{0}\textbf{I})^{-1},
\ena
\]
where $()^{-1}$ and $()^{H}$ refer to inverse and Hermitian transpose, respectively,
$\Phi$ is the inter-cell interference covariance matrix,
and $\textit{N}_{0}$ is the noise power. The transmitted data stream is then estimated using $\textbf{F}^{\rm {MMSE}}$ as
\[
\hat{\bs}[k] = \bF^{\textrm{MMSE}} (\bU_{\textrm{os}}^{\rm H} \bH_{\textrm{os}} \ast \bs[k] + \bw[k])
\]
where $\bw[k]$ is the additive noise terms at time instant $t=kT_s$ ($k = 1,2,...$). 

\section{Simulation Results}
\label{sec:part_Simulation}

In this section, we compare the performance of DPST with that of existing MIMO systems where all the antennas transmit at the same time instant. 

We consider a single tier hexagonal layout of small cell BSs in a $500m \times 500m$ scenario with different ISDs 
to observe the impact of DPST on various degrees of network densification,
which in turn impacts channel correlation. 
The central cell is designated as the serving cell and the remaining six cells are considered as interferers. The carrier frequency and corresponding bandwidth are 2 GHz and 10 MHz, respectively.
Macrocell BSs operate in a different frequency band. 
Each cell consists of an array of two transmit antennas, 
and only serves a single UE (single user scenario), 
which has two antennas forming a $2\times2$ MIMO system. Each signal consists of 10 symbols with an oversampling ratio of 4. Antenna gain, path loss, lognormal shadowing and multipath Rician fast fading are included in SINR computation. The path loss model used is the microcell urban model defined in~\cite{3gpp}, 
which includes both the LOS and NLOS components. 
Closed-loop precoding is considered. Note that it is assumed that the channel is prone to single LOS component.

With regard to DPST, we assume that second transmit antenna is subject to the deterministic delay of $5$ nano sec with respect to first one. Note that the optimization of the precise amount of delay is left as part of future study. We also compare the performance of DPST with respect to the optimistic channel with condition number ($\mathcal{K}$) equals to 1.
This is the optimal channel condition for spatial multiplexing, implying that all channels are orthogonal.
 
\begin{figure}
\centering
\includegraphics[scale=0.53]{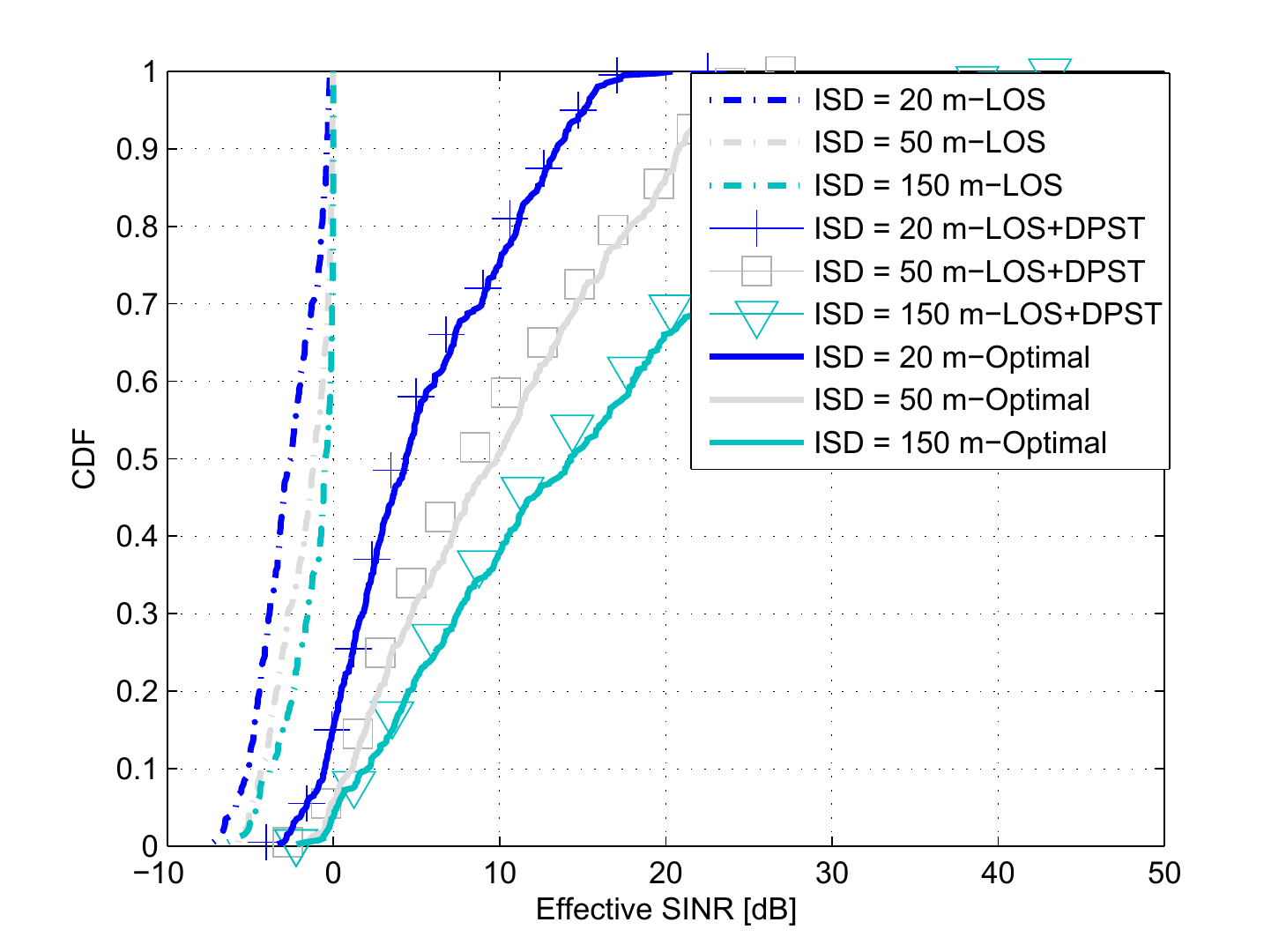}
\caption{Comparison of effective SINR CDF of LOS, DPST and optimum channels in a 2x2 MIMO at different ISDs.}
\label{fig:MIMO2}
\vspace{-0.5cm}
\end{figure}

\begin{figure}
\centering
\includegraphics[scale=0.53]{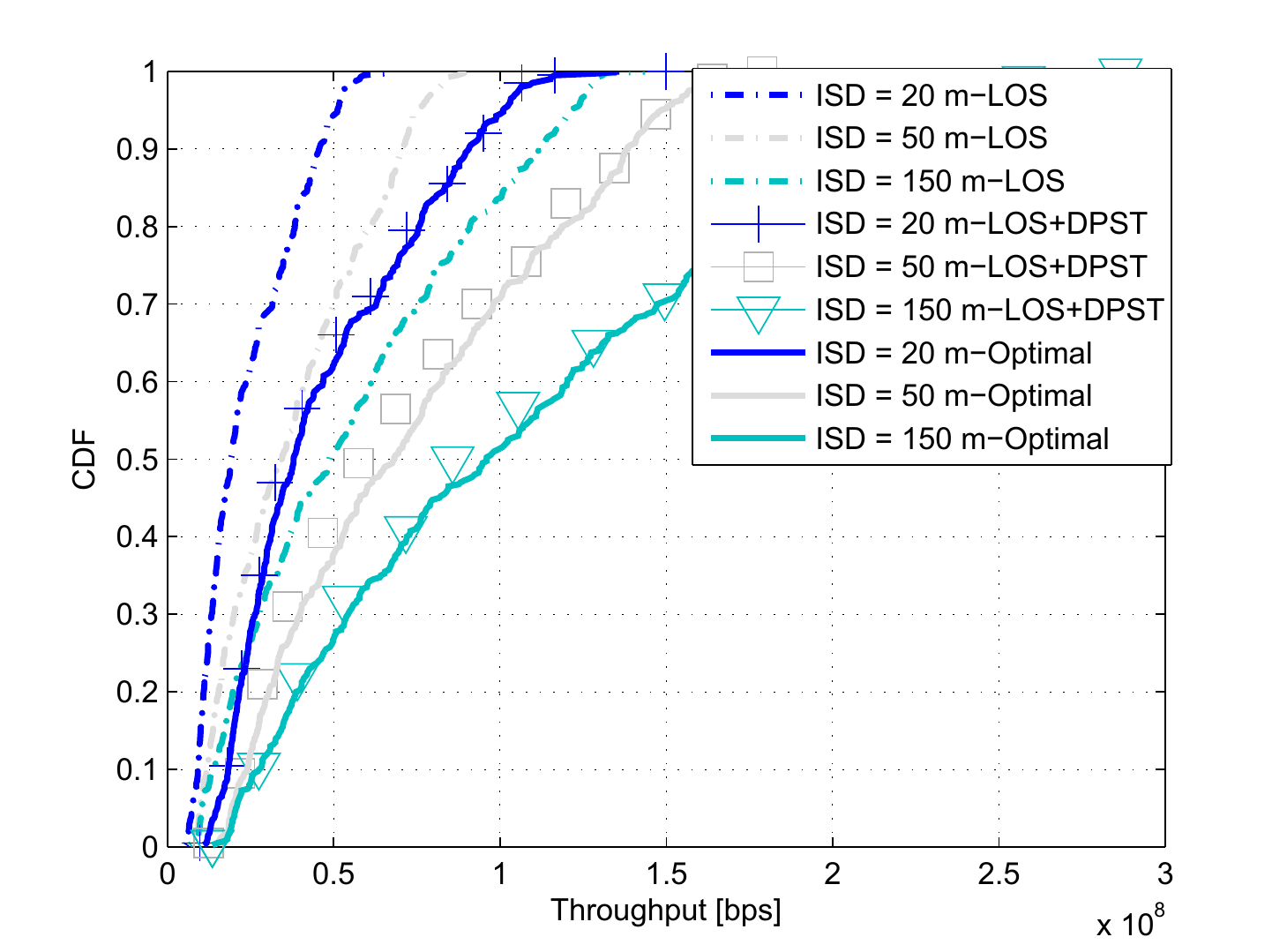}
\caption{Comparison of throughput CDF of LOS, DPST and optimum channels in a 2x2 MIMO at different ISDs.}
\label{fig:MIMO3}
\vspace{-0.5cm}
\end{figure}

Fig.~\ref{fig:MIMO2} compares the effective SINR CDF for a $2\times2$ MIMO system where DPST is applied to the single tap LOS channel model (full correlation). 
Results show that in all tested ISDs, DPST considerably improves the effective SINR with respect to the fully correlated LOS channel, offering a close to optimal performance. 
In more detail, for ISDs of 20~m, 50~m and 150~m, 
the 50\%-tile effective SINR is increased with respect to the LOS  scenario by 6.5~dB, 11~dB and 14~dB, respectively. 
Performance from the optimum is less than 0.1~dB away.
Similarly, Fig.~\ref{fig:MIMO3} shows the throughput CDF of a $2\times2$ MIMO system when DPST is applied to the LOS channel.
For all tested ISDs, DPST enhances the throughput CDF by nearly 2x reaching its upper limit bound defined by optimistic channel condition. This implies that despite the presence of spatial correlation, applying DPST allows the two data streams to be simultaneously transmitted and successfully decoded by the UE receiver. This is in contrary to the correlated LOS channel scenario where the transmitted stream by second antenna can not be decoded by the UE receiver. Table \ref{tab:channel_cond} presents the rank and condition number of the virtual channel when DPST is applied. It is realized that DPST can significantly enhance the virtual channel condition number, verifying its almost optimum performance.

\begin{table}[tbph]
\centering
\renewcommand{\arraystretch}{1.4}
\caption{Comparison of rank and condition number under different channel conditions}
\label{tab:channel_cond}
\scalebox{0.85}{
\centering
\begin{tabular}{|>{\centering\arraybackslash}p{3.4cm}|>{\centering\arraybackslash}p{1.7cm}|>{\centering\arraybackslash}p{1.7cm}|}
\hline \textbf{Channel Status} & \textbf{Channel Rank} & \textbf{Condition Number} \\ \hline
\hline
2 $\times$ 2 LOS Channel & 1 & $\infty$     \\ \hline
2 $\times$ 2 DPST Model & 2  & 1.17 \\ \hline
2 $\times$ 2 Optimum Channel & 2  & 1 \\ \hline
\end{tabular}}
\vspace{-0.5cm}
\end{table}

\section{Conclusion}
\label{sec:part_Conclusion}
 
In this paper, we discussed that in ultra-dense small cell networks, the presence of spatial channel correlation prevents us from achieving spatial multiplexing gain. 
We proposed a new technique referred to as diversity pulse shaped transmission (DPST), 
which exploits pulse shaping diversity in combination with delaying the transmission of one antenna with respect to another by a fraction of symbol period followed by a receiver that operates in the oversampled domain. 
The combined effect is able to virtually lower the channel correlation, improve the channel diversity from the perspective of the UE receiver. 
We studied the performance of the proposed technique under MMSE criteria and compared it with existing $2\times2$ MIMO cellular networks. 
We showed that DPST can significantly enhance the 50\%-tile effective SINR by 11~dB and almost double the UE throughput at an ISD of 50~m. We realize that there is a limit on the number of antennas that benefit from DPST processing which depends on delay optimisation, and hence optimising the corresponding delay as a function of the number of antennas shapes the scope of future research.

\bibliographystyle{ieeetr}
\bibliography{references}

\begin{thebibliography}{10}

\bibitem{2015Lopez}
D.~Lopez-Perez, M.~Ding, H.~Claussen, and A.~H. Jafari, ``{Towards 1 Gbps/UE in
  cellular systems: understanding ultra-Dense small cell deployment},'' in {\em
  IEEE Communications Surveys and Tutorials}, Jun. 2015.

\bibitem{1532224}
J.~Akhtar and D.~Gesbert, ``Spatial multiplexing over correlated mimo channels
  with a closed-form precoder,'' {\em Wireless Communications, IEEE
  Transactions on}, vol.~4, pp.~2400--2409, Sept 2005.

\bibitem{LTE}
F.~Khan, {\em {LTE for 4G Mobile Broadband Air Interface Technologies and
  Performance}}.
\newblock University Cambridge Press, 2009.

\bibitem{6479673}
J.~Anderson, F.~Rusek, and V.~Owall, ``Faster-than-nyquist signaling,'' {\em
  Proceedings of the IEEE}, vol.~101, pp.~1817--1830, Aug 2013.

\bibitem{4777625}
F.~Rusek and J.~Anderson, ``Constrained capacities for faster-than-nyquist
  signaling,'' {\em Information Theory, IEEE Transactions on}, vol.~55,
  pp.~764--775, Feb 2009.

\bibitem{1231648}
A.~Liveris and C.~Georghiades, ``Exploiting faster-than-nyquist signaling,''
  {\em Communications, IEEE Transactions on}, vol.~51, pp.~1502--1511, Sept
  2003.

\bibitem{489269}
J.~Treichler, I.~Fijalkow, and C.~Johnson, ``Fractionally spaced equalizers,''
  {\em Signal Processing Magazine, IEEE}, vol.~13, pp.~65--81, May 1996.

\bibitem{4656965}
S.~Plass, A.~Dammann, and S.~Sand, ``An overview of cyclic delay diversity and
  its applications,'' in {\em Vehicular Technology Conference, 2008. VTC
  2008-Fall. IEEE 68th}, pp.~1--5, Sept 2008.

\bibitem{989781}
W.~Lee, ``The most spectrum-efficient duplexing system: Cdd,'' {\em
  Communications Magazine, IEEE}, vol.~40, pp.~163--166, Mar 2002.

\bibitem{4109390}
S.~Plass and A.~Dammann, ``Cellular cyclic delay diversity for next generation
  mobile systems,'' in {\em Vehicular Technology Conference, 2006. VTC
  2006-Fall. IEEE 64th}, pp.~1--5, Sept 2006.

\bibitem{1192168}
D.~Gesbert, M.~Shafi, D.-S. Shiu, P.~Smith, and A.~Naguib, ``From theory to
  practice: an overview of mimo space-time coded wireless systems,'' {\em
  Selected Areas in Communications, IEEE Journal on}, vol.~21, pp.~281--302,
  Apr 2003.

\bibitem{1203167}
M.~Ivrlc, W.~Utschick, and J.~Nossek, ``Fading correlations in wireless mimo
  communicatin systems,'' {\em Selected Areas in Communications, IEEE Journal
  on}, vol.~21, pp.~819--828, June 2003.

\bibitem{1459054}
A.~M. Tulino, A.~Lozano, and S.~Verdu, ``Impact of antenna correlation on the
  capacity of multiantenna channels,'' {\em IEEE Transactions on Information
  Theory}, vol.~51, pp.~2491--2509, July 2005.

\bibitem{892194}
D.~Chizhik, F.~Rashid-Farrokhi, J.~Ling, and A.~Lozano, ``Effect of antenna
  separation on the capacity of blast in correlated channels,'' {\em IEEE
  Communications Letters}, vol.~4, pp.~337--339, Nov 2000.

\bibitem{2015Jafari}
A.~H. Jafari, D.~Lopez-Perez, M.~Ding, and J.~Zhang, ``{Study on Scheduling
  Techniques for Ultra Dense Small Cell Networks},'' in {\em IEEE Vehicular
  Technology Conference (VTC Fall)}, Sep. 2015.

\bibitem{Matrix}
R.~B. Bapat, S.~J. Kirkland, K.~M. Prasad, and S.~Puntanen, {\em {Combinatorial
  Matrix Theory and Generalized Inverses of Matrices}}.
\newblock Springer, 2013.

\bibitem{5706377}
V.~Venkateswaran and A.-J. van~der Veen, ``Multichannel {S}igma{D}elta {ADCs}
  with integrated feedback beamformers to cancel interfering communication
  signals,'' {\em Signal Processing, IEEE Transactions on}, vol.~59,
  pp.~2211--2222, May 2011.

\bibitem{3gpp}
``{3{GPP} TSG RAN, TR 25.996 v10.0.0, ``Spatial Channel Model for Multiple
  Input Multiple Output (MIMO) simulations (release 10)},'' Mar. 2011.

\end{thebibliography}

\end{document}